\documentclass[11pt]{emulateapj}  
\usepackage{graphicx,amssymb,amsmath,times}
\usepackage{natbib,graphicx}
\usepackage[letterpaper]{hyperref}
\slugcomment{The Astrophysical Journal}
\newcommand{\pks}{\object{PKS~1830$-$211}}
\newcommand{\angstrom}{\mbox{\ensuremath{\mathrm{~\AA}}}}

\newcommand{\keV}{\mbox{\ensuremath{\mathrm{~keV}}}}

\newcommand{\MeV}{\mbox{\ensuremath{\mathrm{~MeV}}}}

\newcommand{\fluxph}{\ensuremath{\mathrm{~ph~cm^{-2}~s^{-1}} }}

\def\latflux{ph cm$^{-2}$ s$^{-1}~$}

\newcommand{\fermi}{\emph{Fermi}~}

\def\swift{\emph{Swift}~}
\begin{document}
%
\title{Gamma-ray flaring activity from the gravitationally lensed blazar\\ PKS~1830-211 observed by {\em Fermi} LAT}
%
\slugcomment{Accepted by The Astrophysical Journal}
%
\shorttitle{Gamma-ray flaring activity of PKS 1830-211 observed by {\em Fermi} LAT}
\shortauthors{Abdo et~al. (the {\em Fermi} LAT Collaboration)}
%
%
%
%

\author{
\footnotesize{
A.~A.~Abdo\altaffilmark{1},
M.~Ackermann\altaffilmark{2},
M.~Ajello\altaffilmark{3},
A.~Allafort\altaffilmark{4},
M.~A.~Amin\altaffilmark{5,62},
L.~Baldini\altaffilmark{6},
G.~Barbiellini\altaffilmark{7,8},
D.~Bastieri\altaffilmark{9,10},
K.~Bechtol\altaffilmark{4},
R.~Bellazzini\altaffilmark{6}
% ...
} 
\vspace{0.2cm}
} 
% ...
\altaffiltext{1}{Center for Earth Observing and Space Research, College of Science, George Mason University, Fairfax, VA 22030, resident at Naval Research Laboratory, Washington, DC 20375, USA}
\altaffiltext{2}{Deutsches Elektronen Synchrotron DESY, D-15738 Zeuthen, Germany}
\altaffiltext{3}{Space Sciences Laboratory, 7 Gauss Way, University of California, Berkeley, CA 94720-7450, USA}
\altaffiltext{4}{W. W. Hansen Experimental Physics Laboratory, Kavli Institute for Particle Astrophysics and Cosmology, Department of Physics and SLAC National Accelerator Laboratory, Stanford University, Stanford, CA 94305, USA}
\altaffiltext{5}{Kavli Institute for Cosmology and Institute of Astronomy, University of Cambridge, Madingley Road, Cambridge CB3 0HA, UK, }
\altaffiltext{6}{Istituto Nazionale di Fisica Nucleare, Sezione di Pisa, I-56127 Pisa, Italy}
\altaffiltext{7}{Istituto Nazionale di Fisica Nucleare, Sezione di Trieste, I-34127 Trieste, Italy}
\altaffiltext{8}{Dipartimento di Fisica, Universit\`a di Trieste, I-34127 Trieste, Italy}
\altaffiltext{9}{Istituto Nazionale di Fisica Nucleare, Sezione di Padova, I-35131 Padova, Italy}
\altaffiltext{10}{Dipartimento di Fisica e Astronomia ``G. Galilei'', Universit\`a di Padova, I-35131 Padova, Italy}
% ...
\altaffiltext{62}{Department of Physics, Massachusetts Institute of Technology, Cambridge, Massachusetts 02138, USA}
%
%
%
%
%
%
%
%
%
%
%
%
%
\normalsize
%
\begin{abstract}
%
%
The Large Area Telescope (LAT) on board the \emph{Fermi} Gamma-ray Space Telescope routinely detects the MeV-peaked flat spectrum radio quasar PKS~1830$-$211
($z=2.507$). Its apparent isotropic $\gamma$-ray luminosity ($E>100$ MeV) averaged
over $\sim$\ 3 years of observations and peaking on 2010 October 14/15 at $2.9 \times10^{50}$ erg s$^{-1}$, makes it among the brightest high-redshift \emph{Fermi} blazars. No published model with a single lens can account for all of the observed characteristics of this complex system. Based on radio observations, one expects time delayed variability to
follow about 25 days after a primary flare, with flux about a factor
1.5 less. Two large $\gamma$-ray flares of PKS~1830$-$211 have been detected by the LAT in the considered period and no substantial evidence for such a delayed activity was found. This allows us to place a lower limit of about 6 on the $\gamma$ rays flux ratio between the two lensed images. \emph{Swift} XRT observations from a dedicated Target of Opportunity program indicate a hard spectrum and with no significant correlation of X-ray flux with the $\gamma$-ray variability. The spectral energy distribution can be modeled with inverse Compton scattering of thermal photons from the dusty torus. The implications of the LAT data in terms of variability, the lack of evident delayed flare events, and different radio and $\gamma$-ray flux ratios are discussed. Microlensing effects, absorption, size and location of the emitting regions, the complex mass distribution of the system, an energy-dependent inner structure of the source, and flux suppression by the lens galaxy for one image path may be considered as hypotheses for understanding our results.
\end{abstract}

%
%
%
\section{Introduction}\label{sect:introduction} %
%
%
The flat spectrum radio quasar (FSRQ) \pks\ (also known as TXS$~$1830$-$210, RX$~$J1833.6$-$210,
MRC$~$1830$-$211, 2FGL$~$J1833.6$-$2104) has met with considerable attention, because it is
such a good example of a gravitationally lensed source. The two lines of sight towards \pks\ have been used as a cosmological probe: temperature of the cosmic microwave background, variations in the fundamental constants, the Hubble constant estimation \citep{bagdonaite13,blandford92}.
This object also offers a unique opportunity to study both the interstellar medium of lens galaxy and the relativistic jet of the background $\gamma$-ray blazar.
\pks\ was
discovered as a single source in the Parkes catalog, but later radio observations by the Very Large
Array (VLA) and Australian Telescope Compact Array (ATCA) clearly
revealed two sources, one in the northeast (NE) and one in the
southwest (SW), separated by 0.98\arcsec\, and connected by an
Einstein ring \citep{pramesh88,Jauncey91}. When the source, lensing foreground object, and
observer lie along a straight line, the theory of gravitational
lensing \citep[e.g. ][]{einstein36} shows that a circle, known as the Einstein ring, may be formed
\citep{schneider92}, while smaller rings could appear inside this main ring if the lens is a Schwarzschild black hole.  The lens magnification factor is the ratio of the flux of the lens image to the flux of the unlensed source, and is equal to the ratio of the
solid angles of the image, and the unlensed source. The NE image has a radio flux density about 1.5 times as bright as the SW one at 8.6 GHz \citep{lovell98}.
Molecular absorption lines revealed
lensing galaxies located at $z=0.88582$ \citep{wiklind96,lovell96,frye99,lehar00,muller11,aller12}
and $z=0.19$ \citep{lovell96} suggesting that \pks\ may be a compound gravitationally lensed system \citep{lovell96}.  These lensing
galaxies were confirmed by Gemini and \emph{Hubble Space Telescope}
(\emph{HST}) \citep{courbin02}. A detailed exploration of this system at
optical wavelengths is hampered by its proximity on the sky to the
Galactic plane and the bulge of the Milky Way (the Galactic
coordinates of \pks\ being $l=12.17$\arcdeg, $b=-5.71$\arcdeg),
leading to considerable dust extinction
\citep[][and references therein]{courbin98,gregg02} and absorption.
Absorption by molecular species ($>30$ different species) from the two foreground galaxies
also peculiarly characterize the radio \pks\ \citep[][]{wiklind96,wiklind98,muller11}.
Molecular absorptions from the intervening galaxy at $z=0.886$, also allowed to put a limit on proton-to-electron mass ratio \citep{bagdonaite13}.

Despite its position near the Galactic plane and center, progress has been made in studying the source in the
optical and near infrared (IR). \citet{courbin98,courbin02} and
\citet{frye99} used a deconvolution algorithm to create
optical/near-IR images of the region, and found the counterparts to
the radio sources, including highly reddened images of the lensing
galaxies.  IR spectroscopy allowed for the redshift of the quasar
itself ($z=2.507$) to be directly measured \citep{lidman99}.

However, even before the redshifts of \pks\ or its lensing galaxies
were known, attempts were made to model the source as a lens
\citep{kochanek92,nair93}.  Since photons for the
source and the image take different paths to reach Earth, it is
expected that there will be a light travel time difference and
consequently a time delay between the photons that arrive from the
different lensed images. That is, variations in the light curve of the SW source
will have the same shape as those from the NE source but arrive later
with a constant time delay and have a smaller magnitude, with respect
to variations in the NE source. Assuming the
same emission region at different frequency bands the
time delay should be the same since
strong gravitational lensing (macrolensing) is an achromatic
process.  Because \pks\ is a blazar that has shown variability
in MeV-GeV bands (COMPTEL, EGRET, AGILE, \fermi)
this opened up the possibility that this time delay can be measured in
$\gamma$ rays.

Modeling combined with redshift and time delay measures can be used to measure Hubble's constant \citep{blandford92}. On the other hand \pks\ is a
compound lensing system, with possible microlensing/millilensing substructures besides the two foreground lensing galaxies at $z=0.886$ and $z=0.19$. Microlensing in the X-ray band is suggested by \citet{oshima01}. An energy-dependent flux ratio of the \pks\ lens images is found in sub-mm bands as clearly associated with the $\gamma$-ray flare of June 2012 and varying with time \citep{ciprini12,marti13}.

A time delay of $\Delta t=26^{+4}_{-5}$\ days was measured from the
light curves of the two lensed images by \citet{lovell98} with
ATCA. They used the values of the delay obtained, along with
the model of \citet{nair93}, to measure Hubble's constant to be $H_0=
69^{+16}_{-9}$\ km s$^{-1}$\ Mpc$^{-1}$, which is consistent with the
most recent measurements \citep[][]{planck_cosmological_parameters_paper}.
Using molecular absorption features,
\citet{wiklind01} found a time delay of $24^{+5}_{-4}$\ days,
consistent with the value found by \citet{lovell98}.  More detailed
modeling of the lensing system, using the time delay of $\Delta
t\approx 25$\ days find similar values of $H_0$
\citep[e.g.,][]{lehar00,witt00}. A different time delay of $\Delta
t=44\pm9$\ days was measured from the radio light curves of the two
lensed images by \citet{vanommen95} using the VLA. \citet{lovell98}
attribute the difference between their measured time delay and the one
found by \citet{vanommen95} as being caused by ``not correctly
accounting for the contribution of the Einstein ring flux density when
calculating the magnification ratio''.

\begin{figure*}[tttt!!]
\resizebox{\hsize}{!}{\rotatebox[]{90}{\includegraphics{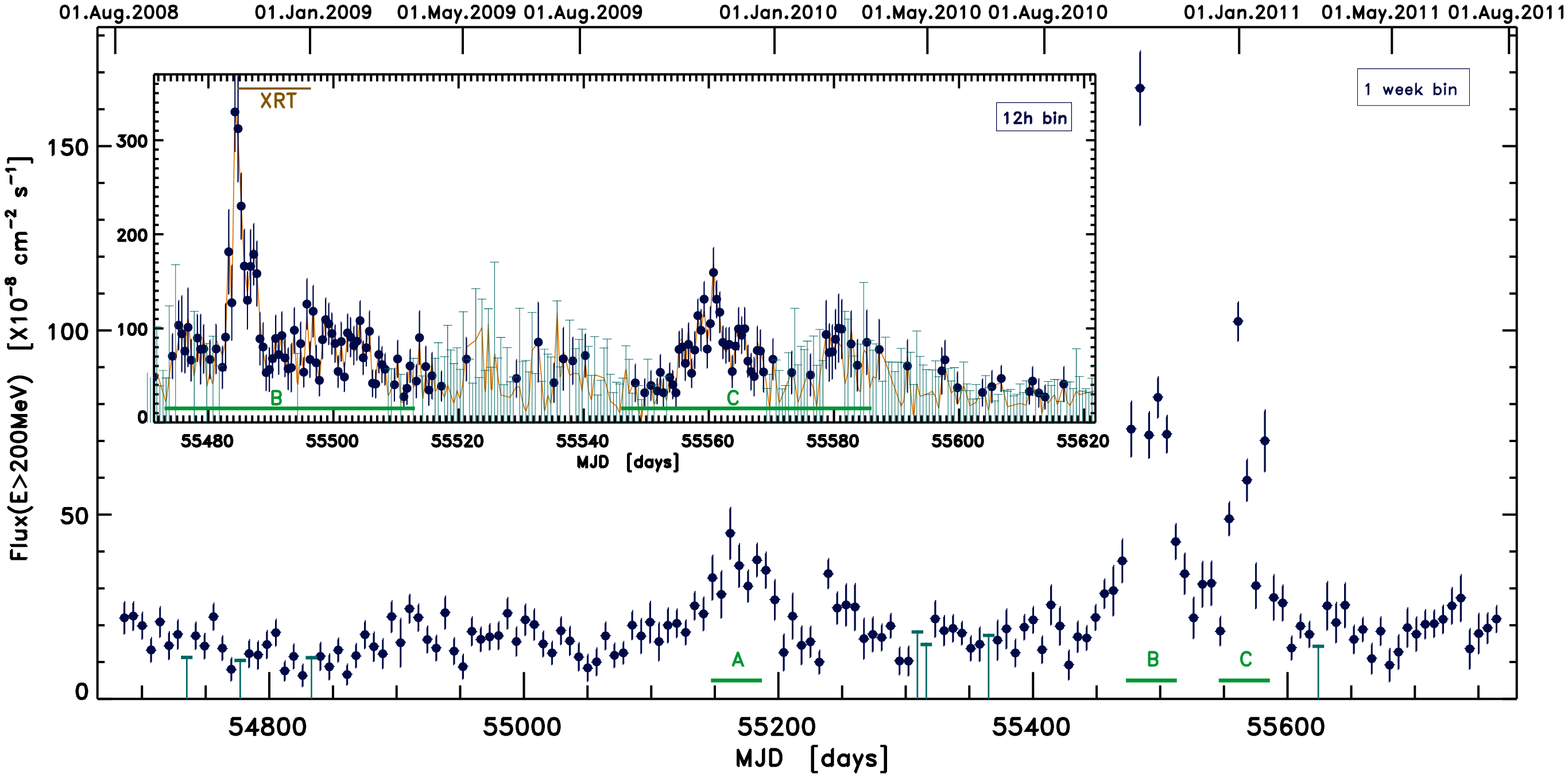}}}\\[-4.5cm]
%
\resizebox{\hsize}{!}{\rotatebox[]{90}{\includegraphics{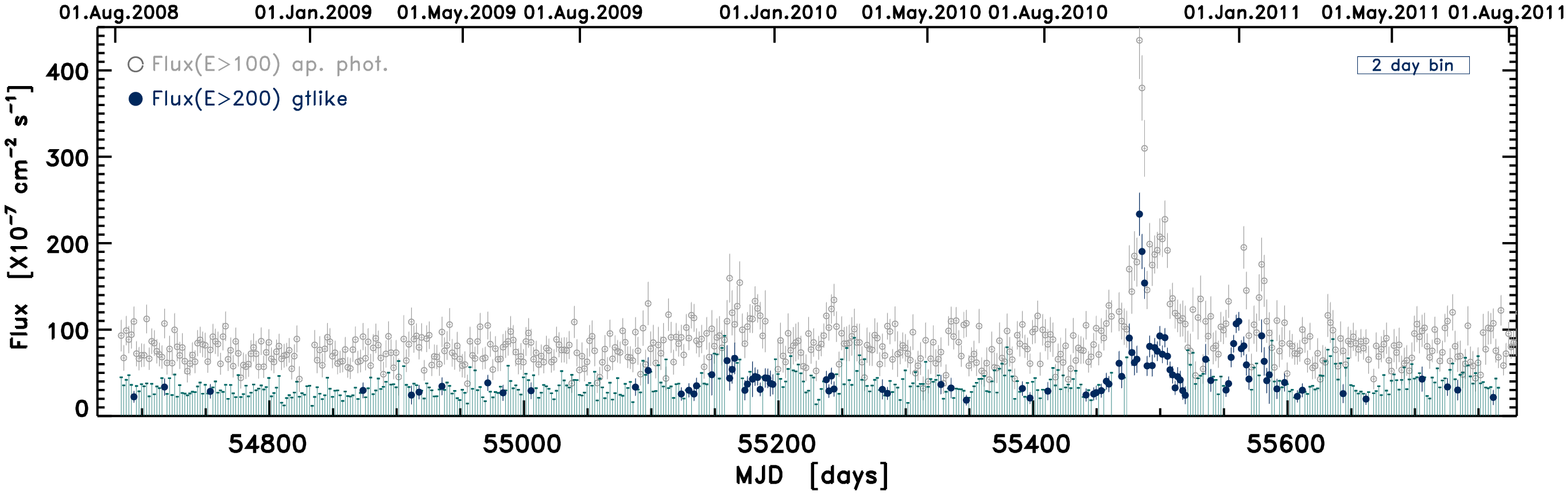}}}
\vspace{-6.3cm}
\caption{
\textit{Top panel}: 3-year (1085 days) LAT $\gamma$-ray flux ($E>200$ MeV) light curve
of PKS~1830$-$211 in weekly bins, extracted with \texttt{gtlike} fit in each bin from 2008 August 4 to 2011 July 25
(MJD 54682.65 to 55767.65).  \textit{Top inset panel}: \texttt{gtlike} light curve
detailing the $\sim 150$ day period (MJD interval: 55471-55621,
i.e. from 2010 October 2 to 2011, March 1) flux light curve extracted
with 12-hour bins and containing the ``B'' and ``C'' intervals when
the main outburst of 2010 October and the second largest, and
double-peaked, flare of 2010 December and 2011 January occurred. In
both panels vertical lines refer to 2-$\sigma$ upper limits on the
source flux. Upper limits have been computed for bins where $TS<4$,
$Npred<3$, or $\Delta F_{\gamma}> F_{\gamma}/2$.
\textit{Lower panel}: 1$\degr$ aperture photometry flux ($E>100$ MeV) and \texttt{gtlike} flux ($E>200$ MeV) light curve of PKS~1830$-$211 in 2-day bins for comparison.
}\label{fig:LCplot}
\end{figure*}
%
%

\pks\ is the brightest gravitational lens in the sky at centimeter wavelengths,
hard X-ray and MeV energies. {\em Chandra}, XMM {\em Newton}, {\em Swift} BAT and INTEGRAL have measured very hard spectra ($\Gamma_{\rm X}\sim1$) and high absorbing column densities accounting for a spectral break below $\sim 4$\keV \citep{derosa05,foschini06,zhang08}.

\pks\ was detected by COMPTEL \citep{collmar06} in the 0.75$-$30 MeV band,
by EGRET \citep[above 100 MeV, 3EG~J1832$-$2110;][]{mattox97a,combi98,hartman99,torres03} and more recently by AGILE \citep[][ and references therein]{striani09,donnarumma11}.

It can be found in the first and second \fermi\ LAT source catalogs
\citep[1FGL$~$J1833.6$-$2103,\\ 2FGL$~$J1833.6$-$2104,][]{1FGLcatalog,LAT_2FGL} with formal significances of about 41$\sigma$ and 67$\sigma$, respectively. The radio source \pks\ and the intervening galaxies are within the LAT error ellipse, as a few nearby field galaxies; nevertheless, there is no source other than \pks\ with radio flux density $\ga10$\ mJy, making it the source of $\gamma$ rays.

Although the NE and SW images of \pks\ cannot be resolved by the LAT, the emission from the two images in principle can be distinguished by measuring a time delay from variable $\gamma$-ray light curves. This
possibility was studied by \citet{barnacka11}, who reported a $ 27.1 \pm 0.6 $ day time delay found in the LAT light curve of this source. This value is in agreement with values found in the radio band \citep[e.g.  ][]{lovell98,vanommen95,wiklind01}.

\pks\ is an FSRQ characterized by a marked $\gamma$-ray Compton luminosity dominance \citep[EGRET/COMPTEL observations, ][]{collmar06}.
The broadband $\nu F_{\nu}$ spectral energy distribution (SED) has been modeled with a combination of synchrotron, synchrotron self-Compton (SSC), and external Compton (EC) scattering of dust torus photons assuming the broadband data were magnified by a factor of 10 by the lens \citep{derosa05}.
\citet{foschini06} and \citet{celotti08} modeled this source without correcting the SED data for extinction or magnification, which are not well known, and used the broad-line region (BLR) as the main seed photon source. Both models provide reasonable descriptions of this object.
Hadronic models predict neutrino production coincident with $\gamma$-ray flares, and this motivates searches for neutrino events coincident with LAT flares \citep[e.g. ][]{icecube13conpks1830}.

\par An outburst observed from the $\gamma$-ray point source positionally consistent with \pks\ was observed by \fermi\ LAT in 2010 October \citep{ciprini10}. This is the largest flare observed since the beginning of the \fermi\ survey, and triggered
rapid-response target of opportunity (ToO) observations by the {\em Swift} satellite\footnote{thanks to a Guest Investigator program {\em Swift} Cycle AO-6 for flaring LAT blazars (PI: L.~Reyes)}.  AGILE also reported a high flux measurement obtained from 2010 October 15 through 17 \citep[][and references therein]{donnarumma11}.

\par In this paper, we explore the $\gamma$-ray properties of \pks\ as observed by the \fermi\ LAT, with particular attention paid to the main outburst of 2010 October and the
second brightest flaring period (2010 December - 2011 January, Section \ref{sect:lat}). In Section \ref{sect:variability} we discuss the $\gamma$-ray flux light curve and the search for time-lag signatures, an indicator of gravitational lensing. The \swift\ observations and results are presented in Section \ref{sect:swift} and the multifrequency SEDs and spectral modeling are reported in Section \ref{sect:sed}. We conclude in Section \ref{sect:conclusion}.

%
%
\section{\emph{Fermi} LAT observations}\label{sect:lat}
%
%
The \fermi\ LAT analysis was performed with the standard LAT
\texttt{ScienceTools} software package (version v9r23p1) and was based
on data collected in the period from 2008 August 4 to 2011 July 25
(from MJD 54682.65 to 55767.65, almost 3 years).  We
first produced a LAT spectrum for \pks\ over this entire time
interval, using only the event class designated as P6\_DIFFUSE (class==3), with corresponding P6\_V3\_DIFFUSE instrument response functions \citep[IRFs, ][]{LATflightperformances} \footnote{These event class and IRFs were used to better compare our results with those of \citet{barnacka11}, where a 2-day bin aperture photometry light curve of \pks\ (300 MeV - 30 GeV flux) was extracted from LAT data with P6\_V3\_DIFFUSE IRFs from 2008 August 4 through 2010 October 13.}, and selecting events in a circular Region of Interest (RoI) with $7\arcdeg$ radius centered on the target position from the second \fermi\ source catalog 2FGL catalog. To reduce contamination from diffuse Galactic emission and nearby point sources, a low-energy cut of 200 MeV was used (compared to the usual cut of 100 MeV where the PSF is relatively large).  A high-energy cut of 100 GeV was also implemented.
\begin{figure}[ttb!!]
\centering 
\resizebox{\hsize}{!}{\rotatebox[]{0}{\includegraphics{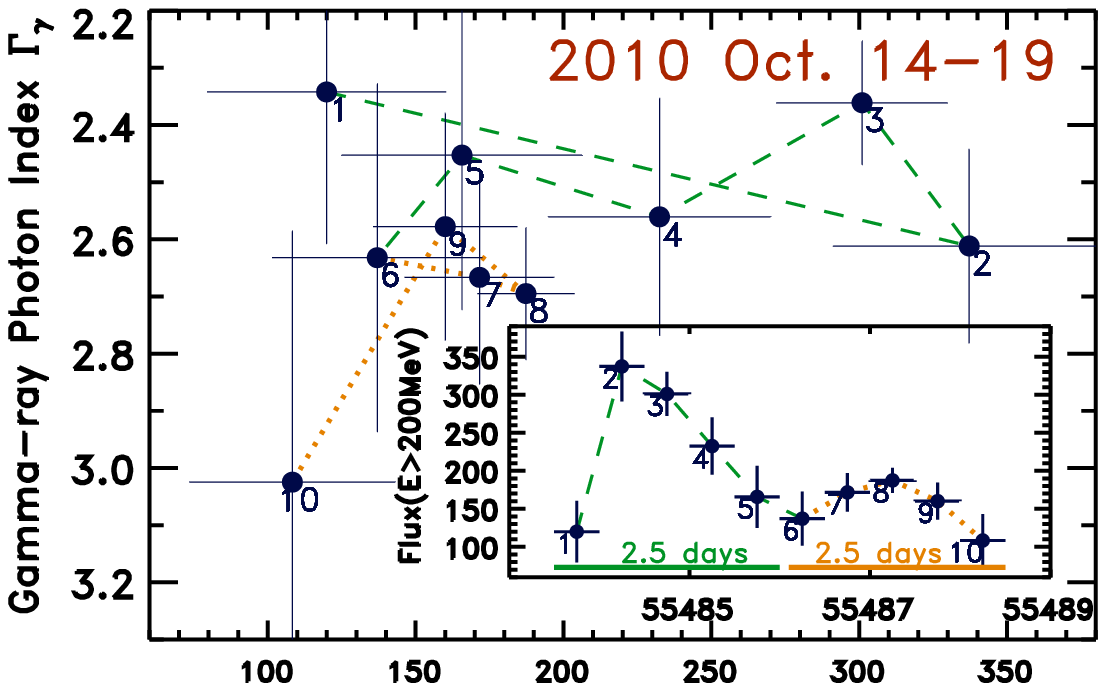}}}\\
\vspace{-0.9cm}
\resizebox{\hsize}{!}{\rotatebox[]{0}{\includegraphics{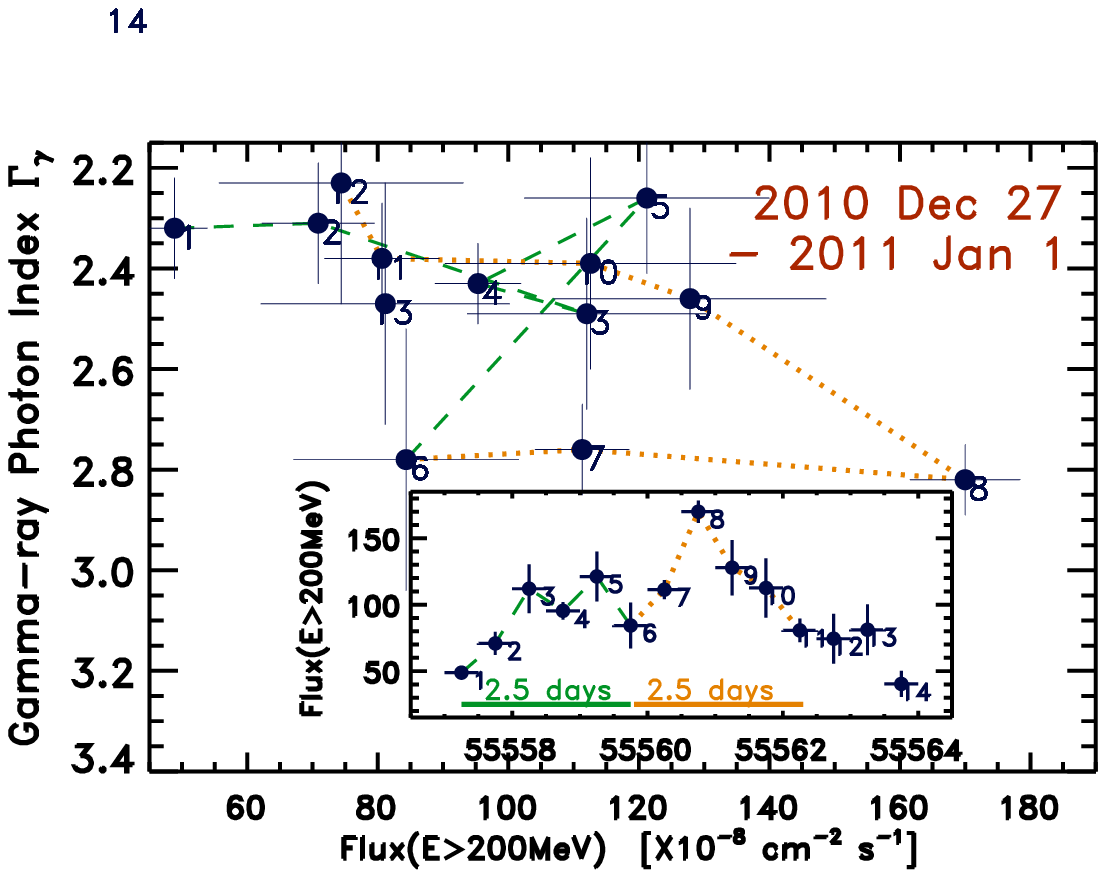}}}
\vspace{-0.4cm} \caption{
5-day zooms on the evolution of the $\gamma$-ray photon index of PKS~1830$-$211 as a
function of the $\gamma$-ray flux during the highest flux peaks of the two main flare events for the source (top and bottom panels). These peaks are contained in the ``B'' and ``C'' intervals of the 12-hour bin light curve reported in top inset panel of Figure \ref{fig:LCplot}. Here the photon index parameter is left free in the likelihood fit. Bars represent $1\sigma$ errors.
} \label{fig:photindexloop}
\end{figure}
%
%

%
\par To avoid the contamination due to the $\gamma$-ray bright Earth
limb in our analysis, all events with zenith angle $>$ 105$^\circ$ were
excluded. Time intervals when the Earth entered
the LAT Field of View (FoV) are excluded selecting only photons with spacecraft rocking
angle $<$ 52$^\circ$.  The unbinned maximum likelihood technique
(\texttt{gtlike} tool) accounted for all the 21 neighboring sources and the
diffuse emission in the physical model of the RoI together with the
target source. The RoI model is fit to the data assuming for the
source \pks\ a power-law spectrum between minimum and maximum energies
($E_{min}$ and $E_{max}$ respectively), $dN/dE\,\propto \,
E^{-\Gamma_{\gamma}}$ with $\gamma$-ray photon index $\Gamma_{\gamma}$
left free in the fit.  A more complex log-parabola model
is reported in the 2-year accumulated data of the 2FGL catalog
\citep{LAT_2FGL} for \pks, but for the purposes of our study and the extraction
of flux light curves in much shorter time bins, the
power-law shape is found to reproduce adequately the source spectrum
(the spectral parameter values obtained with the different models
agree within the statistical errors and the difference in flux values
is found to be on average 5\%).

Source positions were fixed. The Galactic (\texttt{gll\_iem\_v02.fit}) and the isotropic
\\(\texttt{isotropic\_iem\_v02.fit}) background models\footnote{
LAT background models:\\  \hspace{-1.5cm}  \texttt{\scriptsize{fermi.gsfc.nasa.gov/ssc/data/access/lat/BackgroundModels.html}}
} were used with their normalizations left as free parameters in each time bin, facilitating reliable convergence of the likelihood model fits and a reduced computational time.
This procedure was the same as in previous works \citep[e.g., ][]{tanaka13}.
The isotropic component
included both the contribution from the extragalactic diffuse emission
and from the residual charged-particle backgrounds.  In addition, all
$\gamma$-ray sources up to $10\arcdeg$ around the target were included
in the fit with power-law spectral models. The normalization and the photon index
were left free for each point source within a 5$\arcdeg$ radius of
\pks.  Sources between 5$\arcdeg$ and 7$\arcdeg$ had just their
normalizations free (using for each source the fixed photon index
reported in the 2FGL catalog), while sources within 7$\arcdeg$ and
10$\arcdeg$ had all parameters fixed. By exception, the pulsar PSR$~$J1809$-$2332 was modeled with an exponentially cutoff power-law in
which the photon index at low energy, the cutoff energy, and the
normalization factor were left free.
The power-law fit to \pks\ over
the entire period in the $0.2-100$ GeV energy range gave an integrated
flux of $(20.4 \pm 0.4) \times$10$^{-8}$ \latflux and a steep
$\gamma$-ray photon index of $\Gamma_{\gamma} = 2.55 \pm 0.02 $.

Next, we produced a flux light curve for \pks\ using a bin-by-bin maximum likelihood fit (\texttt{gtlike} tool) in
the 200 MeV - 100 GeV energy range with regular time intervals (12 hours, 2 days and 1 week).
We did this assuming the simplest appropriate model, the power-law spectrum,
by freezing the photon index for this source in the individual
time bins equal to the value obtained for the spectral fit over the entire time range,
$\Gamma_{\gamma}=2.55$. This simplification reduced flux error bars produced by the fit minimization
process. Figure \ref{fig:LCplot} (top panel) shows the weekly
(7-day bins) $\gamma$-ray light curve for about the first 3 years
 of the \fermi\ all sky survey.
In the inset panel a 12-hour bin light curve was produced with the likelihood analysis for the 150-day long time interval going from  2010 October 2 (MJD 55471) to 2011 March
1 (MJD 55621) containing the main period of activity for the source with the outburst of 2010 October
and the second largest flare of 2010 December and 2011 January. The lower panel of Figure \ref{fig:LCplot} shows the 2-day bin likelihood flux light curve, mostly characterized by flux upper limits outside the periods of source activity.
For these light curves upper limits have been computed for bins where $TS<4$, the number of
events predicted by the model, $N_{pred}<3$, or the error on the flux,
$\Delta F_{\gamma}> F_{\gamma}/2$. Here $TS$ is the test statistic,
defined as $TS=2\Delta\log(\rm{Likelihood})$ between models with the
additional source at a specified location and without an additional
source, i.e., the ``null hypothesis'' \citep{mattox96}.

We further calculated flux (100 MeV to 200 GeV) light curves in regular time bins (12 hours and 2 days) using the aperture photometry technique \citep[see, e.g., ][]{hadasch12} for the $\sim$ 3-year range (Figure \ref{fig:LCplot}, lower panel, light gray symbols). These flux estimations are extracted using the \texttt{gtbin} tool with an aperture radius of  1$\degr$, and are exposure corrected through the \texttt{gtexposure} tool assuming the power-law spectral shape and the same cuts to photon events reported above. The aperture photometry light curves include a rough background subtraction but, due to the large PSF of the LAT and the nature of the diffuse $\gamma$-ray emission, significant background contamination can be expected as can be seen by the higher ``quiescent'' level with respect to peak of the flares and the level of fluctuations.

The statistical treatment of the likelihood analysis performed in each time bin is the more rigorous approach to extract LAT light curves because of the complications related to the LAT instrument's energy-dependent PSF, geometry-dependent effective area, the nature of the $\gamma$-ray sky backgrounds, the all-sky survey operation mode, and the limited detection rate characterizing GeV $\gamma$-ray data.
\texttt{gtlike} flux light curves provide greater sensitivity and lead to more accurate flux measurements as backgrounds can be modeled out and detailed spectral models can be applied. However, exposure corrected aperture photometry is a useful method for comparisons. It is model independent and more efficient, using short time bins through fewer analysis steps and reduced computational time.

We explore the $\gamma$-ray variability properties further in Section \ref{sect:variability} (and Figures \ref{fig:LCplot} - \ref{fig:wavelet}) with particular attention to possible lensing signatures.
%
%
%

%
\section{Gamma-ray time variability and lensing properties}\label{sect:variability} %

Our analysis is based on the maximum likelihood flux light curves (\texttt{gtlike} tool) while aperture photometry was used to produce supplemental light curves for comparison with the former. In Figure \ref{fig:LCplot} we show the maximum likelihood flux ($E>200$ MeV)
light curve of \pks\ in regular weekly bins over the first 3 years of \fermi\ operation (2008 August 4 to 2011 July 25, MJD 54682.65 to 55767.65). Where $TS<4$, $Npred<3$, or $\Delta F_{\gamma}> F_{\gamma}/2$, 2-$\sigma$ upper limits on the source flux were computed.
1$\degr$ aperture photometry flux ($E>100$ MeV) and likelihood flux ($E>200$ MeV) light curves, both with 2-day bins, are reported (bottom panel of Figure \ref{fig:LCplot}). A likelihood flux light curve in 12-hour bins is also extracted for the $\sim 150$-day interval of the most active phase for the source (``B'' and ``C'' intervals, upper inset panel). Aperture photometry 3-year light curves (1-day/2-day bins) are also extracted in different positions using non-\pks\ photons within the ROI, both along and outside the ecliptic path to better understand possible spurious effects caused by the Sun and Moon passages. We used the same event class and IRFs (P6\_V6\_DIFFUSE) used in \citet{barnacka11} with more checks using P6\_V11\_DIFFUSE IRFs, but
different energy range selection (200 MeV - 100 GeV) and different variability analysis and time bin sizes.

%
\begin{figure}[tttt!!]
\centering 
\resizebox{\hsize}{!}{\rotatebox[]{0}{\includegraphics{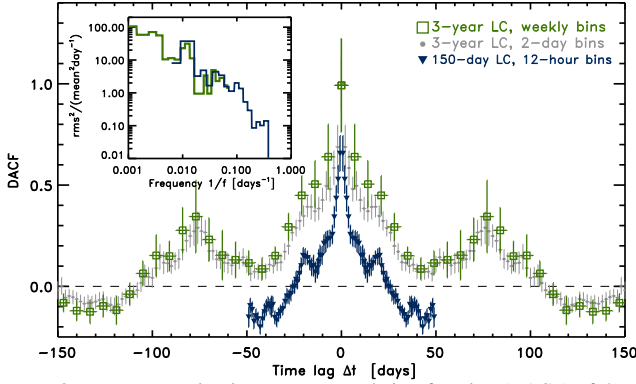}}}\\[-0.3cm]
\hspace{-1.4cm}
\vspace{-0.3cm} \caption{\textit{Main panel}: Discrete autocorrelation function (DACF) of the
  3-year, weekly bin (green square open points) 2-day bin (tiny light gray points)
  and 150-day, 12-hour bin (blue small triangles) LAT flux light curves shown in Figure \ref{fig:LCplot}.
\textit{Inset panel}: Power density spectra, PDS,
  normalized to fractional variance per frequency unit $f$ calculated for the 3-year weekly and the 150-day, 12-hour bin LAT light curves.
} \label{fig:pds}
\end{figure}

\par The ``A'' interval contains the first $\gamma$-ray brightening seen by the LAT, near the end of 2009. The announcement of a detection by AGILE on
2009 October 12 and 13, MJD 55116-55117 \citep[][ and references therein]{donnarumma11} occurred already some weeks before the ``A'' interval.

To explore the behavior of \pks\ during the main outburst (interval ``B'') and the second brightest flaring period (interval ``C'') in greater detail, we performed power-law fits to the source in 12-hour
bins, with both the  flux and photon indices ($\Gamma_{\gamma}$)
left as free parameters.  This is in contrast to the likelihood light curves in
Figure \ref{fig:LCplot} where $\Gamma_{\gamma}$ was fixed.  Note that
12 hours corresponds to $\sim 8$ {\em Fermi} orbits, so that exposures
from bin to bin are roughly the same.  The results can be found in
Figure \ref{fig:photindexloop}, where we searched possible spectral trends.
%
%
\begin{figure*}[tthb!!]
\centering 
\hspace{-0.6cm}
\resizebox{6.3cm}{!}{\rotatebox[]{90}{\includegraphics{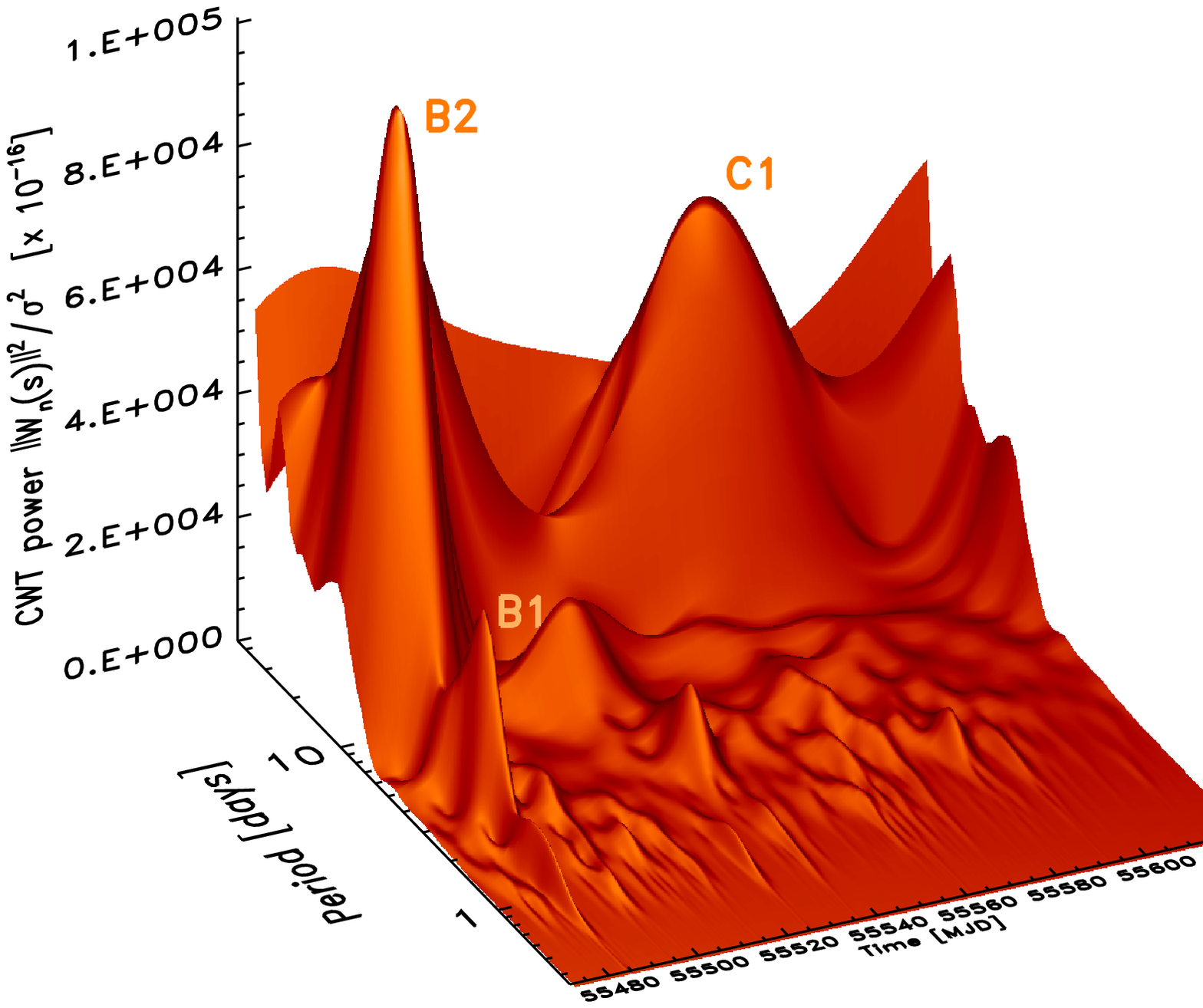}}}
\resizebox{10.5cm}{!}{\rotatebox[]{0}{\includegraphics{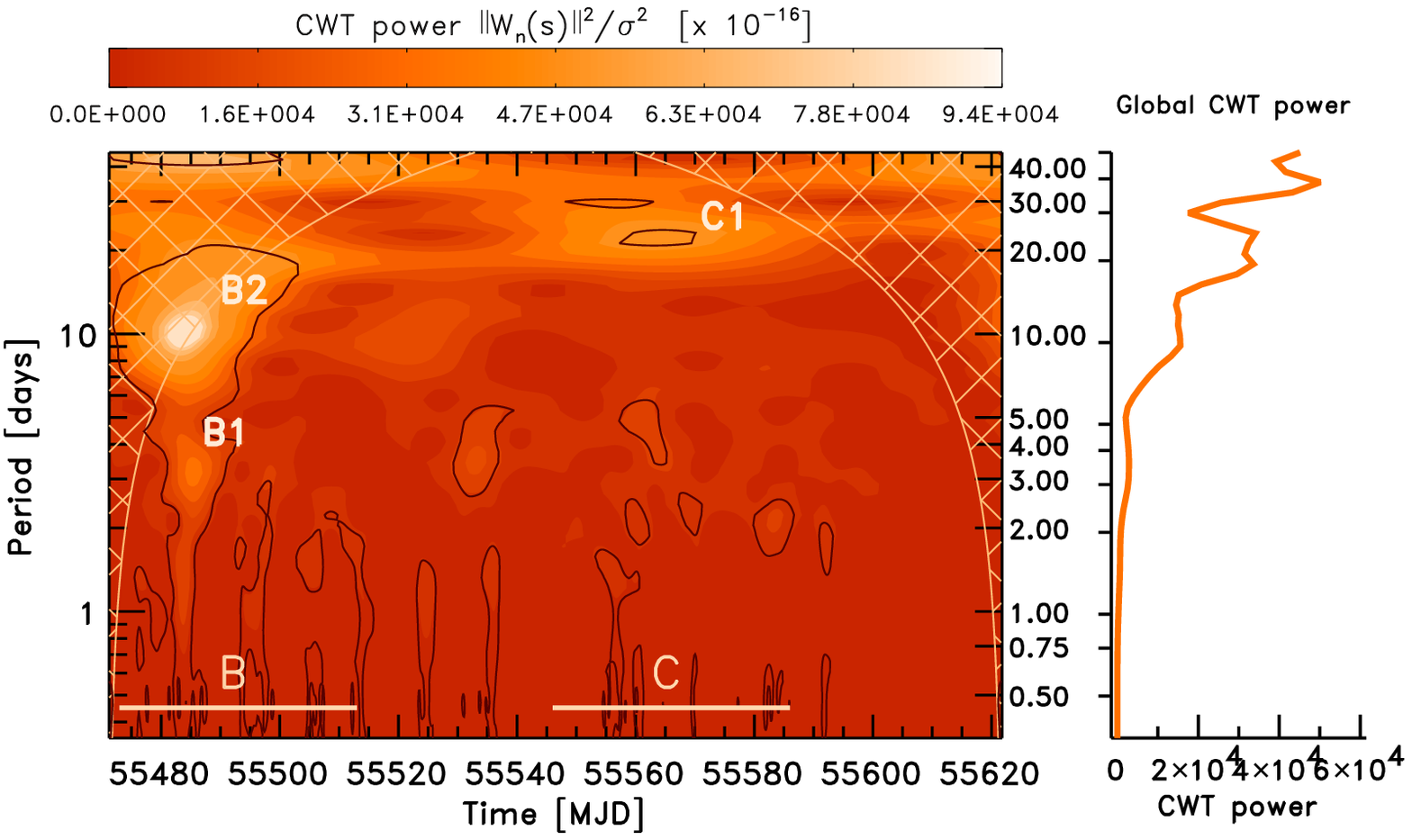}}}
\vspace{-0.2cm} \caption{\textit{Left panel}: plane contour plot of the
continuous wavelet transform power density spectrum (2D PDS from CWT)
for the 150-day and 12-hour bin light curve of Figure \ref{fig:LCplot}
(inset panel), obtained using a Morlet, complex valued, mother
function. Filled color contour plot is the 2D energy density function
of the CWT scalogram. Thick black line contours represents the 90\%
confidence levels of true signal features against white and red noise
backgrounds, while cross-hatched regions represent the ``cone of
influence'', where effects caused by finite time-series
edges become important. \textit{Right panel}: the average of the CWT scalogram over all times is reported, this consists in a smoothed time-averaged 1D CWT spectrum that is called global wavelet spectrum.
} \label{fig:wavelet}
\end{figure*}

The top of Figure \ref{fig:photindexloop} shows the two largest peaks
of structured outburst (within the ``B'' interval) of 2010 October. This is
characterized by a rapid increase of a factor of $\sim2.6$ in flux in
12 hours between 2010 October 14 and 15 (MJD 55483 and 55484) peak of
$F(>200\ \MeV) = (330 \pm 42)\times 10^{-8}$\ \fluxph\ in $\sim 12$
hours, yet taking $\sim$ 48 hours to fall, resulting in an asymmetric
temporal shape. The total peak pulse lasts $\sim 2.5$\ days, and seems to be
followed by another weaker peak also lasting $\sim 2.5$\ days.  Both
peaks do not show significant rotation in the $\Gamma_{\gamma}$-flux
hysteresis diagram, because of the statistically constant photon index
and relatively large uncertainties on flux and $\Gamma_{\gamma}$ with
respect to the variations.

The hysteresis diagram for flare ``C'',
which occurred between about 2010 December 25 and 2011 January 6 (MJD
55555 - 55567) is seen in the bottom of Figure
\ref{fig:photindexloop}.  This flare displays a temporal structure
characterized by two peaks of about 2.5 days duration each. The second peak reaches
a flux value of $(159
\pm 27)\times 10^{-8}$ \fluxph, roughly half of the peak flux of the
``B'' flare.  The flare softens
significantly (bins 6, 7, and 8 in the bottom of Figure
\ref{fig:photindexloop}) to $\Gamma_{\gamma}\sim 2.8$, before turning to its
typical spectrum of $\Gamma_{\gamma}\sim 2.4$ during the decay.

Variability on timescales ranging from about two months down to a couple of days is
seen in these LAT $\gamma$-ray light curves. Two power density spectra (PDS) normalized to fractional variance per unit frequency ($f=1/t$)
(\mbox{\,rms$^{2}$~I$^{-2}$~Day$^{-1}$}), are shown in the inset panel of Figure
\ref{fig:pds}. One is calculated from the 3-year and weekly light curve and one from the
12-hour bin light curve extracted for 150 days between 2010 October 2 (MJD 55471) and 2011 March 1 (MJD 55621) (top main and inset panels of Figure \ref{fig:LCplot} respectively).
Following \citet{lbas-variability}, we consider time bins with flux
upper limits replaced by a value (10$^{-12}$ \latflux), i.e. below
the LAT detection limits. This allows us to
evenly sample the light curve and limit the bias caused
by data gaps. The fraction of upper limits was 5\% and 9\% for the weekly and 12-hour bin light curves, respectively.
Different choices (e.g. replacing upper limits with their half-limit values),
affect the PDS slope estimates by a few percent, which is substantially less than other uncertainties. The white
noise level was estimated from the rms of the flux errors and was subtracted for each PDS.

Both PDS are in good agreement with each
other, meaning that the fractional variability and its time scale
distribution during the more active ``B'' and ``C'' epochs are the same as during the longer and fainter
periods between the flaring events. The merged PDS is fit with a
simple $1/f^{\alpha}$ power law, with a slope $\alpha= 1.25 \pm 0.12$,
while the low-frequency PDS is fit with $\alpha=1.1 \pm 0.2$, and the
high frequency one with $\alpha=1.3 \pm 0.2$.

The main panel of Figure \ref{fig:pds} reports the Discrete
Autocorrelation Function (DACF) for the three likelihood flux light curves reported in Figure \ref{fig:LCplot}: the 3-year weekly and 2-day bin, the 150-day 12-hour bin light curves. The weekly and 2-day bin light curves follow the same function
profile, consistent with the PDS power-law index, showing no signal
power peak for timescales shorter than peak hinted at $76\pm4$
days. The origin of this value could be time series noise or related to the  time between the ``B'' and ``C'' flare peaks. The control aperture photometry light curves for non-\pks\ photons extracted in different positions within the RoI do not provide signals at any timescale.

The DACF of the best-sampled 12-hour bin light curve shows a clearer peak of $19\pm 1$ days that is not evident in the 2-day bin DACF. That value could represent a possible characteristic timescale of the time series, created by a regular gravitational lensing time delay as found in \citet{barnacka11}. This might represent a possible point of rough agreement with their results. On the other hand this DACF peak can be produced by the timescale of the two main flare events (the peaks in the ``B'' and ``C'' intervals, Figures \ref{fig:LCplot} and \ref{fig:photindexloop}). A power spectrum analysis that is time-localized along the light curve epochs, like the wavelet method, can help to shed light on this.
The 150-day duration, 12-hour bin, light curve was also analyzed using a Continuous Wavelet Transform (CWT) analysis (Figure
\ref{fig:wavelet}). By decomposing the light curve into time-frequency ($t$, $f=1/t$) generalized Fourier spaces, we are able to determine both the
dominant modes of variability (as with the PDS), as well as how those
modes vary in time, localizing them along the light curve epochs. This produces a diffuse and continuous two-dimensional
(2D) time-frequency (or time-period) image plot, ``the scalogram'' (Fig. \ref{fig:wavelet}, left and central panels).
In such a plot we report the normalized 2D modulus of the CWT energy density function ($\left\| W_{n}(s) \right\|^{2}/\sigma^{2}$, where
the normalization $1/\sigma^{2}$ gives a measure of the power relative
to white noise), computed using a Morlet mother waveform. This mother function provides the best tradeoff between time-localization and period (frequency) resolution. Thick black line contours are the 90\% confidence levels of true signal features
against white and red noise backgrounds, while cross-hatched regions
represent the ``cone of influence'', where spurious edge effects
caused by finite time-series boundaries become important.

Most of the CWT power, not influenced by edge effects, is concentrated within the period scales (y-axis) ranging
from 8 to 30 days, even if there is appreciable power at longer periods (e.g., at 40-50 days). The main outburst (``B'') is decomposed and resolved in time/frequency (x-y) spaces with the bulk of the power released between
about MJD 55475 and 55495 (2010 October 6 - 26) peaking at MJD 55484
(2010 October 15), in agreement with the light curve shape. The corresponding characteristic scale is
10 days, which is related to the peak duration. The outburst is also
characterized by a resolved timescale component of $\sim 3$ days at MJD
55486 (October 17) in agreement with the 2.5-day peaks substructures mentioned (Fig. \ref{fig:photindexloop}, top panel).
This timescale still appears significant but drifted to longer values
of about 3.5 days and 4.5 days respectively during the events at
around MJD 55535 (2010 December 5) and MJD 55563 (2011 January 2,
i.e. the flare epoch ``C''). This second brightest flare event for \pks\ is identified
by a significant (within 90\% confidence local region) and
well-defined peak of CWT signal power with characteristic timescale
of about 21 days, between about MJD 55560 and 55565 (2010 December 30
and 2011 January 4) in agreement with the previous description and
Figures \ref{fig:LCplot} and \ref{fig:photindexloop} (bottom
panel).

Summarizing, between the main outburst ``B'' and the
second brightest flare ``C'', we observed a shift
from a characteristic timescale of the main outburst of $\sim
10$ days (``B2'' peak in the CWT plot) to a timescale
of $\sim20$ days for the second flare event (``C1'' peak in the CWT plot).
This suggests the ``C'' flare phase has twice the duration of the ``B'' flare phase, yet is approximately half
as bright in emitted $\gamma$-ray power. The CWT analysis implies both these timescales are characteristic of the coherent and separate flare events ``B'' and ``C''.  It does not provide evidence for a detection of a regular signal recurrent along the whole light curve as produced by gravitational lensing.

Based on a time delay of $\Delta t=26^{+4}_{-5}$\ days measured by \citet{lovell98} and
$24^{+5}_{-4}$ days measured by \citet{wiklind01}, the main outburst ``B'',
beginning between 2010 October 14 and 15 (MJD 55483-55484), should
have a delayed event occurring within the time interval MJD
55503 - 55514 (2010 November 3 - 14).  If the delay measurement of
$44\pm 9$ days \citep{vanommen95} is correct, this would put the
$\gamma$-ray flare from the lens image starting around 2010 November
27 (MJD $55527\pm 9$). A delayed $\gamma$-ray flare event would appear in the CWT scalogram
as a clear peak of power, separated on the horizontal (time) axis by the time delay from the ``B'' and the ``C'' flares.
Delayed flares should also be visible in the best-sampled 12-hour bin flux light curve \citep[as found in the LAT light curve of S3$~$0218+35, lens B0218+357][]{cheung13}, contrary to
our findings. The few peaks in the CWT power spectrum at about 27 days after the flaring epochs are not
significant. In any case, the chance coincidence of two flares within 20--30 days has a non-negligible probability. A detection of a feature in the DACF such as the 19-day peak does not provide enough evidence of a detection of delayed events induced by lensing, as testified by the lack of a well-resolved peak in the epoch-averaged (global) CWT power spectrum (third, right panel of Figure \ref{fig:wavelet}).

Exposure-corrected aperture photometry provides supplementary information (Figure \ref{fig:LCplot}), even if the technique is limited by lower sensitivity, inaccuracies in the definition the model for diffuse $\gamma$-ray emission, small aperture area (and so also fewer photons from the source).

The unbinned likelihood potentially offers greater sensitivity, more accurate flux measurements, and reduced uncertainties and fluctuations. Both methods can be affected by spurious instrumental modulations and systematic errors. One possible effect is correlated to the dependence of the particle background rate on the \fermi\ spacecraft geomagnetic location, which is modulated by the orbit precession\footnote{Precession period of $\sim 53.2$ days as reported in \citet{LATflightperformances} and inferred from
NORAD two-line element sets (\texttt{www.celestrak.com/NORAD/elements/}).}. Lunar $\gamma$-ray emission may influence every light curve extraction for \pks\ \citep{corbet13}. The $\sim 27$ day scale can be consistent both with the first harmonic of the \fermi\ spacecraft orbit precession period and the Moon's sidereal period. The comparison aperture photometry light curves extracted from non-\pks\ positions in our data do not show any significant signatures of such effects.

We note that changes in observing conditions and other instrumental effects could induce temporal correlations in measured quantities. Therefore, uncertainties in both the light curve and derived model parameters might be
underestimated owing to the potential for additional low-level sources of systematic temporal correlations.

The absence of clear evidence for delayed flare episodes following the ``B'' and ``C'' events, and the lack of regular time scale signatures in our 3-year LAT data, imply either that lensing delayed flares at $\gamma$-ray energies do not exist in this source, or the flux ratio in the $\gamma$-ray band does not match that observed in the radio bands ($\sim 1.5$). We might also be observing a time-dependent or energy-variable lensing flux ratio. A varying ratio (range 1.0 - 1.8) of the measured flux of the two radio images is already suggested in \citet{wiklind98}. Multi-year monitoring of the absorption caused by the $z=0.886$ galaxy showed temporal changes in absorption lines, ascribed to motion of the blazar images with respect to the foreground galaxy and produced by sporadic ejection of bright plasmons \citep{muller08}.
We discuss our result further in Section \ref{sect:nodelay}.

%
%
\section{\emph{Swift}: data analysis and results}\label{sect:swift}
%
%

The {\em Swift} satellite \citep{gehrels04} performed 10 ToO observations on
\pks\ between 2010 October 15 (16:26 UT) and October 27 (09:07 UT), MJD 55484.685-55496.380, for a GI program
triggered by the high $\gamma$-ray activity of the source. The {\em Swift}
observations were performed with all three on-board instruments: the X-ray
Telescope (XRT, 0.2--10.0 keV), the Ultraviolet/Optical
Telescope (UVOT, 1700--6000 \angstrom), and the Burst Alert Telescope (BAT, 15--150 keV).

\subsection{{\em Swift} BAT observations}

The hard X-ray flux of this source is below the sensitivity of the BAT
instrument for the short exposures of the \swift\ ToO observations
performed on 2010 October. The source was not detected between 2010
October 14 and 18 (net exposure of about 200 ks) by INTEGRAL
\citep{donnarumma11}. By contrast, \pks\ is detected in the BAT
58-month catalog, generated from the all-sky survey from 2004 November
to 2009 August. Therefore, we used the 8-channel spectrum  available
at the HEASARC\footnote{\texttt{http://heasarc.gsfc.nasa.gov/docs/swift/results/bat}}. The 14--195 keV spectrum is well described by a power law with photon index of 1.50$\pm$0.13
($\chi^2_{red}$/d.o.f. = 0.89/6). The resulting unabsorbed 14-195 keV
flux is (9.0$\pm$0.8)$\times$10$^{-11}$ erg cm$^{-2}$ s$^{-1}$. The difference in flux and photon index between the 58- and 70-month BAT catalog spectra is negligible.

%

\subsection{{\em Swift} XRT observations}

The XRT data were processed with standard procedures ({\tt xrtpipeline} v0.12.4), including the filtering, and screening criteria from the {\tt Heasoft} package (v.6.8). The source count rate was low during all the observations (count rate $<0.5$ counts s$^{-1}$), so we only considered photon counting (PC) data and further selected XRT event grades 0--12. Source events were extracted from a circular region with a radius between 15 and 25 pixels (1 pixel $\sim 2.36\arcsec$), while background events were extracted from a circular region with radius 40 pixels away from background sources. Ancillary response files were
generated with {\tt xrtmkarf}, and accounted for different extraction regions, vignetting and PSF corrections. We used the the redistribution matrix function version v011 in the Calibration Database maintained by HEASARC. All spectra were rebinned with a minimum of 20 counts per energy bin to allow $\chi^2$ fitting within {\sc XSPEC} (v12.5.1).

Previous soft X-ray observations of \pks\ revealed a hard spectrum ($\Gamma_{\rm X}\sim1$) and absorption in excess of the Galactic column due to the lensing galaxy at $z=0.886$ \citep{mathur97,oshima01,derosa05}. In particular, \citet{derosa05} derived a value of column density for this extra absorption of 1.94$^{+0.28}_{-0.25}$ $\times$10$^{22}$ cm$^{-2}$ from a broad band spectra with {\em Chandra} and INTEGRAL data. {\em XMM-Newton}
observations of \pks\ were modeled by \citet{foschini06} with a broken
power law model, with the photon index changing from $\sim$1.0 to
$\sim$1.3 at about 3.5 keV. The joint fit of XMM/INTEGRAL data
performed by \citet{zhang08} confirmed that the broken power law is
the best model fit, with column density, photon indices and energy
break parameters very similar to those found in the previously-cited
works.

\par
\begin{figure}[tttt!!]
\centering 
\hskip 0cm
\resizebox{\hsize}{!}{\rotatebox[]{0}{\includegraphics{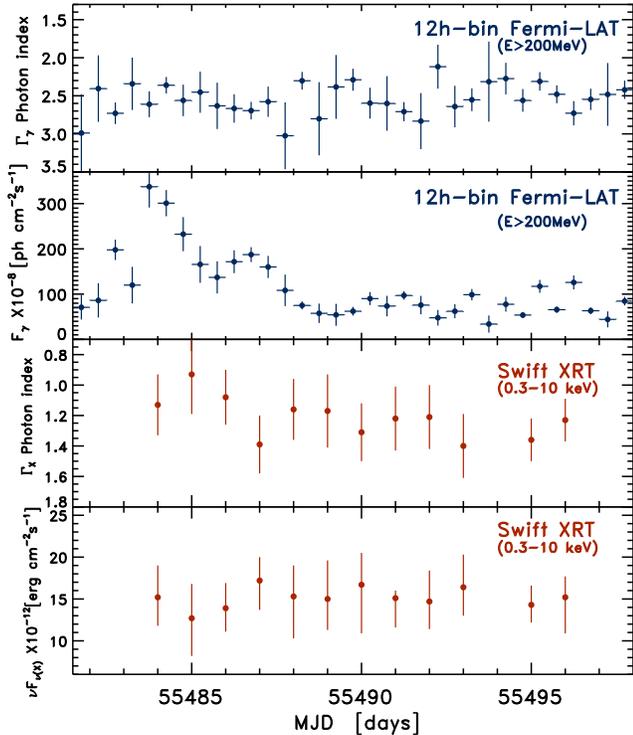}}}
\vspace{-0.2cm} \caption{
Multi-panel plot with simultaneous \fermi\ LAT and \swift\ XRT flux and photon index light curves.
} \label{fig:LAT_and_XRT}
\end{figure}

 We fit the individual XRT spectra of 2010 October with an
absorbed power law, with a neutral hydrogen column fixed to its
Galactic value ($2.05\times10^{21}$ cm$^{-2}$; Kalberla et al.~2005)
and an extra absorption fixed to the value found by De Rosa et
al.~(2005). The resulting unabsorbed 0.3--10 keV fluxes and the photon
indices for each observation are reported in Figure \ref{fig:LAT_and_XRT}. The unabsorbed flux derived from XRT observations lies between 1.3 and 1.7 $\times$ 10$^{-11}$ erg cm$^{-2}$ s$^{-1}$.

%
%

To investigate in more detail the X-ray spectral properties of the
source we accumulated all the events collected during this campaign
for extracting an average spectrum with higher statistics. As a first
step we fit the average spectrum with the same model used for the
single observations, obtaining an acceptable fit. Leaving the value of the column density of the extragalactic absorber free to vary, a comparable fit is recovered, with larger uncertainties on the parameters. We found instead an
improvement in the fit substituting the simple power law with a broken
power law model, significant at the 99.9$\%$ confidence level
according to the F-test.

The 0.3--10 keV flux detected by XRT in 2010 October is
only slightly higher than those observed in the past XMM-\textit{Newton} and \textit{Chandra} observations of the source (Table \ref{tab:swift}), indicating no significant activity in soft X-ray during the LAT flare.

A joint fit to the XRT+BAT spectrum with an absorbed broken power law
and a cross-correlation factor between XRT and BAT of 1.30$^{+0.36}_{-0.28}$ led to a further slight improvement ($\chi^{2}_{red}$/d.o.f. = 1.09/133), with photon indices $\Gamma_{X1}$ = 1.05$\pm$0.10 and $\Gamma_{X2}$ = 1.53$\pm$0.11 below and above a break energy of 3.59$^{+0.83}_{-0.51}$ keV.

\subsection{{\em Swift} UVOT Observations}

During the {\em Swift} pointings, the UVOT instrument
observed \pks\ in the $v$, $b$, $u$, and $uvw1$, $uvm2$ and
$uvw2$ photometric bands.  The analysis was performed using the {\tt
uvotsource} tool to extract counts from a standard 5\arcsec\ radius
aperture centered on the source, correct for coincidence losses,
apply background subtraction and co-add all of the individual images
for each filter. Nevertheless, due to the high extinction in the direction of
\pks, the source was not detected above 3-$\sigma$ in any of
the UVOT bands, so we computed a 3-$\sigma$ flux upper limit (lower
limit in magnitude) for each filter: $v > 18.0$, $b > 19.5$, $u >
19.3$, $uvw1 > 16.9$, $uvm2 > 20.0$, and $uvw2 > 21.0$.

\begin{table}[t!!!!]
\begin{center}
\scriptsize
\vspace{-0.2cm}
\caption{Summary of the \emph{Swift} XRT analysis of the PKS 1830$-$211 ToO observations.\label{tab:swift}}
\begin{tabular}{ccccc}
\tableline\tableline
\multicolumn{5}{c}{Power law Model}\\
\tableline
Exp\tablenotemark{a} & $N_{\rm H}$\tablenotemark{b} & $\Gamma_{X1}$  & Flux (0.3--10)$^{c}$ &  $\chi^{2}_{\rm r}$/(d.o.f.) \\
\tableline
20.3 & 1.94 (fix) & 1.20$\pm0.06$ & $1.64\pm 0.11$ & 1.19\\
 &  &  & & (129) \\
20.3 & 2.09$^{+0.54}_{-0.36}$  & 1.23$^{+0.11}_{-0.08}$ & 1.65$^{+0.27}_{-0.18}$  & 1.19\\
 &   &  &   &  (128) \\
%
%
\tableline
\tableline
\multicolumn{5}{c}{Broken Power law Model}\\
\tableline
Exp\tablenotemark{a} & $N_{\rm H}$\tablenotemark{b} & $\Gamma_{X1}$/$\Gamma_{X2}$\tablenotemark{d} & Flux
(0.3--10)\tablenotemark{c} & $\chi^{2}_{\rm r}$/(d.o.f.) \\
\tableline
20.3 & 1.94 (fix) & 1.05$\pm0.10$ & 1.53$^{+0.14}_{-0.11}$ & 1.13 \\
  &   & 1.56$^{+0.39}_{-0.20}$ &  & (127) \\
\tableline
\end{tabular}
\begin{list}{}{}
(a) Net exposure in kiloseconds adding the single XRT
observations performed between 2010 October 15 and 24. \\   \vspace{0.1cm}
(b) Column density of the extragalactic absorber at redshift
  z=0.886 in units of $10^{22}$~cm$^{-2}$. A Galactic absorption of 2.05 $\times$ $10^{21}$~cm$^{-2}$ (Kalberla et al.~2005) is added.  \\   \vspace{0.1cm}
(c) Unabsorbed flux in the $0.3-10$~keV energy band.  \\   \vspace{0.1cm}
(d) $E_{\rm break}=3.65^{+1.35}_{-0.60}$ keV.
\end{list}
\normalsize
\end{center}
\vspace{-0.1cm}
\end{table}
%

%
\section{Broadband spectral energy distribution}\label{sect:sed}
%

\begin{figure*}[htt!!]
\centering 
\resizebox{15.5cm}{!}{\rotatebox[]{0}{\includegraphics{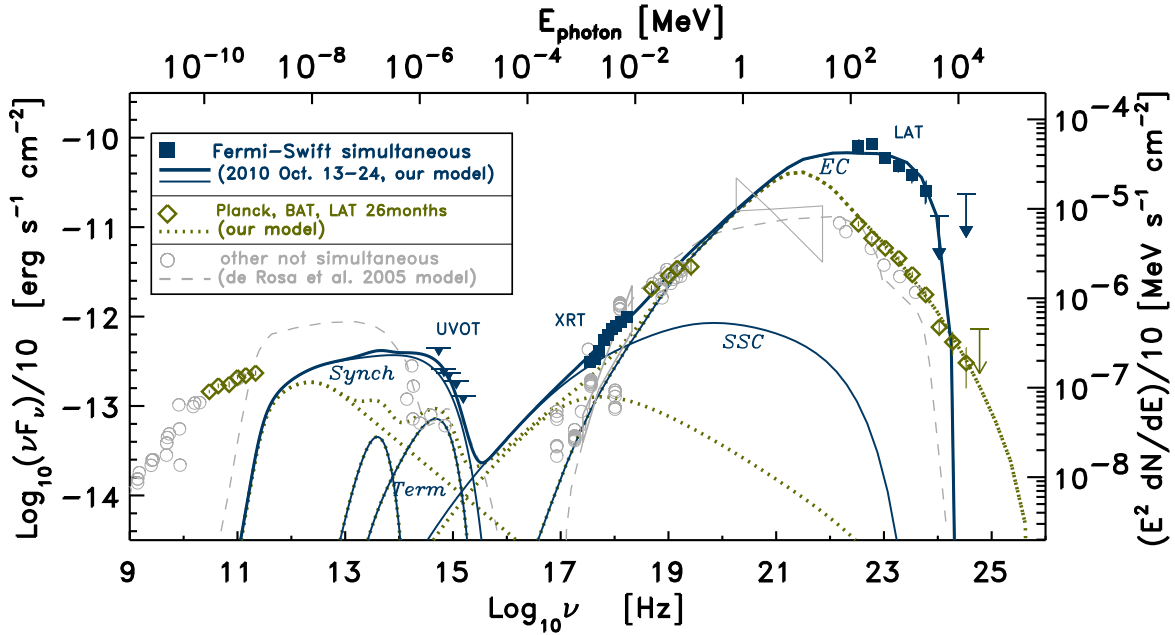}}}
\vspace{-0.2cm} \caption{
The SED of PKS~1830$-$211 built with simultaneous \fermi\ LAT
and {\em Swift} XRT and UVOT (upper limits only) data, averaged over
the 2010 October 13-24 campaign and corresponding to the $\gamma$-ray
outburst (all plotted as blue/dark square symbols). Also plotted are a
non-simultaneous 26-month LAT
spectrum, the BAT 58-month spectrum, and the {\em Planck} ERCSC
spectrum (all plotted as green/dark open diamond symbols).  Archival
data from radio/mm, Gemini-N, {\em HST}, {\em Chandra} (2001 January),
INTEGRAL IBIS (2003), COMPTEL (bowtie), and EGRET are taken from
literature \citep{derosa05,foschini06,zhang08} and are plotted as
light grey open circles with a light gray dashed-line model. All data
are corrected for lensing for a factor of 10 magnification (although note that the
magnification may not be the same for all frequencies; see section
\ref{sect:nodelay}). Also plotted are fits with a synchrotron/SSC/EC
model to the outburst state (blue/dark solid curves fitting the LAT, XRT, UVOT simultaneous campaign data) and to the low activity state (green/dark dotted curves) represented  by LAT, BAT, ERCSC not simultaneous data.
} \label{fig:sed}
\end{figure*}

The $\nu F_{\nu}$ SED of \pks\ around the 2010 October outburst (epoch ``B'') is
shown in Figure \ref{fig:sed}.  The data have been de-magnified by a
factor of 10, following \citet{nair93} and \citet{mathur97}.
Rarely can pure synchrotron/SSC models reproduce the observed
SEDs of FSRQs.  We attempted to fit PKS 1830$-$211 with such a model,
but similar to \citet{derosa05}, we were not able to adequately
reproduce its SED, since the SSC component is too broad to reproduce
the X-ray and $\gamma$-ray data.

The high activity observed in $\gamma$ rays has no significant
counterpart in soft X rays, but those data can be described by a
single EC component, suggesting that the X-ray photons originated in
the low-energy tail of the same electron distribution. To fit the
simultaneous 2010 October SED with an EC model, we assume that the
emitting region is at a considerable distance from the black hole,
outside the BLR, and that the primary seed photon source is from a
dust torus emitting blackbody radiation in the infrared. There is some
debate about the location of the $\gamma$-ray emitting region,
although a large distance from the black hole seems justified for
FSRQs by detailed campaigns by the \fermi\ LAT and radio
observatories \citep[e.g.,][]{marscher10}.  The dust torus was assumed
to be a one-dimensional annulus with radius $r_{dust}$ centered on the
black hole and aligned perpendicular to the jet, and emitting
blackbody radiation with temperature $T_{dust}$ and luminosity
$L_{dust}$. The dust parameters were chosen to be consistent
with the sublimation radius \citep{nenkova08}.  For the disk
luminosity in our models, the BLR radius would be at $2\times10^{17}$\
cm, using the scaling relation of \citet{ghisellini08}.  We place the
emitting region at a distance 10 times that of the BLR, making Compton
scattering of BLR photons negligible \citep{dermer09}.

%
%

Our best fit is shown as the blue curve in
Figure \ref{fig:sed}, and the parameters of the fit are described in
Table \ref{fitparams_table}. The entries in this table are free parameters except the jet powers and the blob radius.  The model and parameters are described
in detail by \citet{dermer09}. The emitting region size scale chosen
is consistent with a variability time scale of 12 hours, observed for
the main outburst epoch (``B'').

We found that an electron distribution with two spectral breaks (three
power laws) was necessary to reproduce the SED.  A very hard $p_1$ was
necessary to fit the hard XRT spectrum.  The other electron indices,
$p_2$ and $p_3$, were chosen to be the same as the fit by
\citet{derosa05}.  Due to the simultaneous non-detection at UV/optical
wavelengths, our model is not strongly constrained.  Notice that the
Compton-scattered peak is $\sim 10^3$ times larger than the
synchrotron peak, and that this is really a lower limit on the
Compton-dominance, due to the lack of an optical detection.  For the
outburst ``B'' SED fit, the total jet power, $P_{j,B} + P_{j,e}\approx
3.3\times10^{45}$\ erg s$^{-1}$ is below the Eddington luminosity for
a $10^9\ M_{\odot}$ black hole ($L_{Edd}\approx1.3\times10^{47}$ erg
s$^{-1}$), as one would expect, and the magnetic field and nonthermal
electrons are within approximately a factor of 4 from equipartition.
Again, as with the fit by \citet{derosa05}, this fit is also able to
explain the X-ray and $\gamma$-ray emission with a single EC
component.  The fit is also similar to the one by \citet{foschini06}.

We also built a ``quiescent state'' SED of \pks\ from non-simultaneous
data: the 58-month BAT spectrum, the {\em Planck}
Early Release Compact Source Catalogue, ERCSC, spectrum, the LAT first-26 month spectrum.
This LAT spectrum excludes the prominent flaring activity in 2010
October and 2010 December/2011 January, and so should be a fairly good
representation of the source during low-variability and low-activity
states (green/dark data points with diamond symbol and dotted green/dark lines fit in Figure \ref{fig:sed}).

We also included the other relevant
archival data (gray open circle data points in Figure \ref{fig:sed}, with instruments
indicated in the caption). The dust and disk emission are the same
for both models.  We found that we could
reproduce the quiescent state SED by varying only two parameters from
the outburst state SED, namely the highest electron index ($p_3=4$) and the
cutoff of the electron distribution ($\gamma_{max}=10^5$).  This attempt
provides a decent fit to the archival data, except for the COMPTEL
bowtie. However, since these are non-simultaneous, this should not be
considered a major deficiency in the modeling. The archival optical emission here comes mainly from the accretion
disk, so that this fit is also poorly constrained.
Finally, the model fit from \citet{derosa05} is shown for comparison.
The model is quite similar to ours, although it provides a bit better
fit to the archival optical and COMPTEL data.

%
\section{Discussion and conclusions}\label{sect:conclusion}
%

We have presented detailed \fermi\ LAT $\gamma$-ray and {\em Swift}
observations of the gravitationally-lensed and MeV-peaked FSRQ \pks.
The LAT analysis was based on data collected in the period from
2008 August 4 to 2011 July 25 (from MJD 54682.65 to 55767.65, about 3
years). Increased $\gamma$-ray activity
of this source was detected in 2009 November followed by a large outburst in
2010 mid-October, namely epoch ``B'', and a
second flare at the period between 2010 December and 2011 January,
namely epoch ``C''.  \pks\ stands out for a number of reasons,
besides the fact that it is characterized by strong-type gravitational lensing, which we discuss further in Section \ref{sect:nodelay}.

\pks\ is the third most distant object detected in large flaring activity so
far by \fermi\ LAT behind \object{TXS~0536$+$145}
and \object{B3~1343$+$451}. The apparent isotropic $\gamma$-ray
luminosity ($E>100$ MeV) of \pks\ over the first 31 months of \fermi\ operation is $\sim$1.1$\times$10$^{49}$ erg s$^{-1}$,
comparable to the brightest high redshift ($z \gtrsim 2$) blazars in
the Second LAT AGN Catalog \citep[][ 2LAC]{lat-2lac}.

\begin{deluxetable*}{lccc}
\vspace{-0.6cm}
\tabletypesize{\scriptsize}
\tablecaption{
Model fit parameters
}
\tablewidth{0pt}
\tablehead{
\colhead{Parameter} &
\colhead{Symbol} &
\colhead{2010 Oct 13--24 fit} &
\colhead{Quiescent fit}
}
\startdata
Bulk Lorentz Factor & $\Gamma$	& 20 & 20	  \\
Doppler Factor & $\delta_D$       & 20 & 20   \\
Magnetic Field & $B$         & 1 G & 1 G  \\
Variability Timescale & $t_v$       & 12 hours & 12 hours \\
Comoving Blob radius & $R_b^\prime$ & $7.4\times10^{15}$\ cm & $7.4\times10^{15}$\ cm \\
Jet Height & $r$ & $10^{18}$\ cm & $10^{18}$\ cm \\
Low-Energy Electron Spectral Index & $p_1$       & 1.0 & 1.0    \\
Medium-Energy Electron Spectral Index & $p_2$       & 1.8 & 1.8    \\
High-Energy Electron Spectral Index  & $p_3$       & 2.8 & 4.0  \\
Minimum Electron Lorentz Factor & $\gamma^\prime_{min}$  & $3$ & $3$ \\
First Break Electron Lorentz Factor & $\gamma^\prime_{brk1}$  & $30$ & $30$ \\
Second Break Electron Lorentz Factor & $\gamma^\prime_{brk2}$  & $300$ & $300$ \\
Maximum Electron Lorentz Factor & $\gamma^\prime_{max}$  & $6\times10^3$ & $1\times10^5$ \\
\hline
Black Hole Mass & $M_{BH}$ & $10^9\ M_{\odot}$ & $10^9\ M_{\odot}$ \\
Accretion Disk Luminosity & $L_{disk}$ & $3.9\times10^{P45}$ erg s$^{-1}$ & $3.9\times10^{45}$ erg s$^{-1}$ \\
Inner disk radius & $R_{in}$ & $6 R_g$ & $6 R_g$ \\
Blob distance from black hole & $r_{blob}$ & $10^{18}$  & $10^{18}$  \\
\hline
Dust torus temperature & $T_{dust}$ & $1.7\times10^3$\ K & $1.7\times10^3$\ K \\
Dust torus radius & $r_{dust}$ & $2\times10^{18}$ cm & $2\times10^{18}$ cm \\
Dust torus luminosity & $L_{dust}$ & $3.1\times10^{45}$\ erg s$^{-1}$ & $3.1\times10^{45}$\ erg s$^{-1}$ \\
\hline
Jet Power in Magnetic Field & $L_{j,B}$ & $1.6\times10^{44}$ erg s$^{-1}$ & $1.6\times10^{44}$ erg s$^{-1}$ \\
Jet Power in Electrons & $L_{j,e}$ & $3.8\times10^{45}$ erg s$^{-1}$ & $3.1\times10^{45}$ erg s$^{-1}$ \\
Total Jet Power & $L_{j,tot}$ & $4.0\times10^{45}$ erg s$^{-1}$ & $3.3\times10^{45}$ erg s$^{-1}$  \\
\hline
\enddata
\label{fitparams_table}
\vspace{-0.2cm}
\end{deluxetable*}

The $\gamma$-ray flux observed by the LAT from this source was at its peak on 2010 October
14-15, reaching a flux of $F(E>200\ \rm{MeV})\ \approx300\times10^{-8}$
ph cm$^{-2}$ s$^{-1}$, as seen in the 12-hour binned light curve.
This is a factor of 17 greater than the average 3-year flux.  The
corresponding apparent isotropic $\gamma$-ray luminosity of
$2.9\times10^{50}$ erg s$^{-1}$ is greater than that observed from
\object{PKS~1622$-$297} during the 1995 flare \citep{mattox97b}, and
from \object{3C~454.3} in 2009 December \citep{3c454}, and roughly
comparable to the 2010 November outburst from this source
\citep{3c454_02}. For this bright flare, if one uses the variability
timescale in the proper frame of the source $\Delta t\approx 12\
\rm{hours}/(1+z)\approx 1.3\times10^4$\ s, and a de-magnified
luminosity of $L_\gamma\approx3\times10^{49}$\ erg s$^{-1}$, one
calculates $L_\gamma/\Delta t \approx 2.5\times10^{45}$\ erg s$^{-2}$.
This value is a bit below the record-holder for AGN, from the 2010
November burst from \object{3C 454.3} \citep{3c454_02}, but it still
exceeds the \citet{elliot74} limit of $L_{Edd}/(R_S/c)\approx
1.3\times10^{43}$\ erg s$^{-2}$ (where $R_S$ is the Schwarzschild
radius), and the limit that includes Klein-Nishina effects,
$1.6\times10^{44}$ erg s$^{-2}$ \citep{liang03}.

No correlated variability for this $\gamma$-ray flare was detected in
X rays by {\em Swift} XRT, which is somewhat typical for FSRQs
\citep[e.g.,][]{marscher10,hayashida12}, although not
universal \citep[e.g.,][]{raiteri11}. Orphan $\gamma$-ray flaring
activity in \pks\ was already found in AGILE data
\citep{donnarumma11}. This fact, in addition to the lack of detection
in optical/UV by {\em Swift} UVOT and hard X-ray by INTEGRAL IBIS
during the 2010 October $\gamma$-ray flare discovered by \fermi\ LAT,
indicates the mechanism producing the $\gamma$-ray flare only
marginally influences the X-ray part of the spectrum.  There may be
correlated variability between $\gamma$-ray and optical emission, also
typical for FSRQs \citep[e.g.,][]{marscher10,raiteri11}, but without
any optical detections, it is impossible to tell. The lack of X-ray
and $\gamma$-ray correlation can support the lack of evident signals
of strong lensing at high energies.

The hard and soft X rays are thought to be a combination of the
contributions from SSC and EC, and the soft X-ray roll-off is
explained in terms of a natural interplay between SSC and EC
components \citep{foschini06}. The extremely hard X-ray photon indices
have been found for a number of other blazars \citep{sikora09}, and
seem to indicate very hard electron distributions at low energies.

The main (``B'') outburst of 2010 October was found asymmetric with a fast
rise of a factor about 2.6 in flux in 12 hours, a phenomenology observed in a few
$\gamma$-ray blazar flares in the past \citep{lbas-variability}.
The asymmetry might imply particle acceleration and cooling times that are greater
than the light crossing time, i.e., $t_{\rm inj}, t_{\rm cool} > R/c$
(in the jet comoving frame).  The fast rise and slower decay shape can also be evidence for a contribution by Comptonization of photons produced outside the jet
\citep{sikora01,sokolov05}.

A 2.5-day flux peak timescale appears to characterize the ``B'' and ``C'' flares.
The main outburst ``B'' was characterized by a 10-day timescale, but there was
a shift to a timescale of about 20 days during the ``C'' flare.  That
is, the timescale of the emission doubled. This is based on the 2D CWT
scalogram for the 12-hour bin light curve, and is
supported by the CWT global spectrum and the DACF (with a peak at $19
\pm 1$ days). This scale appears to be unconnected to a regular
lens-delay signature running along the whole light curve: it is at the
boundary of the range of the radio delay values \citep[$\sim
20-30$][]{lovell98,wiklind01} and would not be well compatible with the
$\sim$ 27-day value found in \citet{barnacka11}. In general
multi-scale variability ranging from months down to a couple
of days is found in the LAT light curves.
In particular the 76-day
interval separating the peaks of the two main flare episodes ``B'' and ``C'', namely
the peak of 2010 October 15 (MJD 55484) and the peak of 2010 December
30 (MJD 55560) emerged as a possible signature from the DACF analysis. In terms of
gravitational lensing this lag cannot be connected to radio-band lag values,
and only one episode is not sufficient to make
further speculations in this direction. The fractional $\gamma$-ray variability and its
timescale distribution during the more active phases are found to be
similar to the the ones shown in the longer, fainter and less variable
intervals between the flare events, and the PDS can be described by a
$1/f^{1.25 \pm 0.12 }$ power law. This implies the occurrence of a
specific variation is inversely proportional to its strength, with
more weight preferred for short timescales.

The steep
$\gamma$-ray spectrum of MeV-peaked sources like \pks\ can contribute
to the cosmic X-ray background and the extragalactic $\gamma$-ray
background, depending by luminosity functions as well as SED models.
The 3-year LAT data analysis presented in this work
suggests its $\gamma$-ray flaring activity and temporal behavior are
due to intrinsic variability within the source, rather than to
strong gravitational lensing effects. Data acquired in the next
years of \fermi\ all-sky survey monitor will shed more light on the meaning of the hinted timescales.

\subsection{Why has no time delay been observed in gamma rays?}\label{sect:nodelay}

The intense $\gamma$-ray flaring from \pks, the brightest LAT
gravitationally lensed blazar together with S3$~$0218+35
\citep[i.e. the lens system B0218+357][]{cheung13}, has opened up the possibility of measuring
$\gamma$-ray time delays from the different lensed images of the
blazar.  The first clear $\gamma$-ray measurement of a delay for the images of the
lens B0218+357 is reported in \citet{cheung13}, where a lag of $\simeq 11.5$ days, that is
$\sim 1$ day greater than previously determined radio-band values, was
determined. Inspecting the intervals around the brightest flares of this source,
magnification flux ratios in $\gamma$-ray energy bands
were measured oscillating about unity, with magnitudes smaller than
those from radio observations.  In the case of \pks\ however, as we
show in Section \ref{sect:variability}, the expected delay of $\approx
25$\ days with a flux ratio $\approx 1.5$ \citep[e.g.,][]{lovell98}
was not found by us with enough evidence, despite first claims to the contrary \citep{barnacka11}. We can set a lower limit of $\sim 6$ for the $\gamma$-ray flux ratio between the two lens images, significantly larger than the flux ratio in radio bands.
The two radio images correspond to very slightly different viewing angles of the background blazar ($\Delta \theta \sim 1 $ arcsec); therefore any source emission anisotropy, such as relativistic beaming, can change the observed flux ratio. This first limit found by us in $\gamma$-rays implies a very small beaming angle for the $\gamma$-ray emission.

\pks\ is a case of both strong lensing (characterized by a
double image) and a compound lensing induced by two foreground
galaxies. For an ideal lens the flux image ratios in different energy
bands should be the same as the deflection is achromatic
\citep{schneider92}. Multiple imaged quasars and blazars are expected to show intrinsic variability in all the resolved lensed images with the same
time delay. Variable differences between the light curves could be
ascribed to microlensing acting on the system. Inhomogeneities and radiation absorption can
significantly change the observed flux and lensing magnification.
In particular some material can interfere with the $\gamma$-rays in the lens galaxy and suppresses those from the SW image of \pks.
In \citet{winn02} the SW image of \pks\ is observed to pass through one of the spiral arms of the $z=0.19$ foreground galaxy.

Different flux ratios have been measured from other lensed quasars
\citep[e.g.,][]{blackburne06,pooley06,chen11}. Those authors
attributed this to microlensing substructure in the lensing system and
a different spatial origin of the emission at different wavelengths
(X-ray and optical emission in those cases).  This has been shown to
be possible through lens modeling \citep{dobler06} and can
explain the observed different flux ratio with respect to the radio one.
The amplitude of the magnification caused by microlensing is greater for small emission regions. Production sites for GeV $\gamma$-rays are generally much smaller than those at radio bands ($<0.003$ pc from our SED modeling). Microlensing in the lens foreground galaxy could therefore produce further flux modulations and variations of the light curve produced by stellar motions in the galaxy. Optical microlensing is observed in some galaxies, and $\gamma$-ray emitting regions are comparable to the optical continuum size of an AGN. Based on EGRET data of \pks\ $\gamma$-ray flux variations are already suggested to be produced by gravitational microlensing \citep{combi98}.
Microlensing could allow to constrain the postulated power-law relationship $R \propto E^{a}$ between size and energy of $\gamma$-ray emission regions, and could explain some of the unidentified LAT $\gamma$-ray sources at high galactic latitude through lensing magnification of background undetected blazars \citep{torres03}.

The typical time scale for a caustic-crossing microlensing event in a lensed quasar however is longer than $\sim 25$ days \citep[weeks, months][]{fluke99,wambsganss01}. On the other hand modeling of microlensing events has also shown that microlensing durations can be
different for different wavelengths when the emission originates from
different size scales \citep{jovanovic08}. In addition \pks\ has rather fast source crossing times and a small ratio of source size to Einstein radius \citep{mosquera11}, therefore significant microlensing variations are expected for this lensed $\gamma$-ray blazar.

The evidence for gravitational microlensing and millilensing effects in strong lensed quasars is increasing in recent works \citep[e.g. ][]{blackburne11,chartas12,chen12}. Microlensing structures or light path time delays sampling intrinsic quasar spectral variability are thought to explain optical spectral differences between quasar image components \citep[e.g.,
][]{wisotzki93,sluse07,sluse13}.
In X-rays spectral variations can be described by changes of absorption column density, and by different spectral components and broken power laws, with different absorptions. It may be possible that the X-ray beam passes through a high-absorption column, but the radio band image is covered by a partial absorber with a low covering factor. In the peculiar case of \pks, energy dependence observed in X-ray flux ratio between the two images is also ascribed to microlensing events \citet{oshima01} inducing time variability and X-ray chromatic perturbations. For example, X-ray microlensing variability was identified and disentangled in the \textit{Einstein Cross}
QSO 2237+0305 \citep{zimmer11}.

The observed flux of resolved lens images $(i)$ at time $t$ is a result of different factors: $F^{(i)}(t)=\mu^{(i)}_{macro} \cdot \mu^{(i)}_{micro} \cdot Q(t) \mu^{(i)}_{macro}$, where $\mu$ are the macro/micro-lensing magnification factors and $Q(t)$ the time-dependent flux of the quasar. The Einstein-ring radius on the \pks\ source plane is
$R_{E}=\theta_{E}D_{os} \simeq 2 \times 10^{16}\sqrt{M_{lens}/M_{\sun}}  $ \citep{paczynski86,oshima01}.

We can use the lower limit of $\sim6$ in
the $\gamma$-ray flux ratio to put an upper limit on
the size of the $\gamma$-ray emitting region
\citep{grieger91,yonehara98}.  We find that this must be $R_b^\prime
\la 5.6\times10^{14}m^{1/2}$\ cm, where $m$ is the mass of a microlens
in solar masses.  This is consistent with the $\gamma$-ray variability
timescale, although it is larger than the size used in SED modeling
(section \ref{sect:sed}). The size of the $\gamma$-ray emission region evaluated from the SED modeling ($7.4\times 10^{15}$ cm) is smaller than $R_{E}$ and is therefore subject to possible microlensing, inducing magnification variations with respect to radio wavelengths where the emission region is more extended. In particular a
larger magnification ratio is expected for a caustic-crossing microlens event \citep{blandford92} acting on one of the two images, as suggested by the $\gtrsim 6$ $\gamma$-ray flux ratio. It should be noted that microlensing due to individual stars in the main lens galaxy is expected to be negligible in many cases, as the projected Einstein radius of each star is smaller than the \pks\ optical source extension. However, further lensing effects can be due to nearby galaxies. There are six other secondary galaxy candidates for weak lensing, identified in the field within 20'' from the main lens by \citet{lehar00}. These galaxies can provide lensing effects exerted
at the position of the NE and the SW images of \pks, in the case that one or more of them is relatively massive and placed at $z \lesssim 0.1$. In this case they would have to be included in \pks\ lens modeling.

Besides micro/milli-lensing effects and the need for a refined strong lensing modeling, there are other open possibilities that could explain the lack of an evident $\gamma$-ray lensing time delay
for the two major flares of \pks\ seen by the LAT. \textit{Chandra} and XMM-\textit{Newton} observations of \pks\ show large variations in the absorption column density, which are interpreted as intrinsic absorption \citep{dai08}.
As the \pks\ X-ray emission is dominated by relativistically
beamed components from the jet, it is very likely that the obscuration may be due to jet-linked absorbing material, physical processes, or variations from the geometric configuration of the jet.
If the $\gamma$-ray emission region is displaced from the radio-band emission region the $\gamma$-ray flux ratio can have the observed difference from the radio flux ratio. Spatially distinct emission regions may give some constraint on differing jet structure probed by the two different energy regimes. This hypothesis may complicate results for this blazar
in comparing the radio/$\gamma$-ray properties, in evaluating Compton dominance, and in correctly modeling its SED. It could be possible that the radio
and $\gamma$-ray emission in blazars comes from different regions of
the jet with different size scales.  This is due to the well-known
fact that variability at these different wavelengths is on
considerably different timescales, and that compact synchrotron
emission from jets is strongly self-absorbed at radio frequencies.
The $\gamma$-ray emission site for the quiescent period from August 2008 to September 2010 and the $\gamma$-ray emitting region responsible for the two main flaring episodes ``B'' and ``C''
also could be different \citep{barnacka14} with different lensing magnification ratios. In general the magnification ratio might differ for radio-band and
$\gamma$-ray emission, especially when there are high-energy flaring episodes.

Multi-epoch and multi-frequency continuum observations of the two
resolved lensed images of \pks\ by ALMA in the 350-1050 GHz band showed a remarkable
frequency-dependent behavior of the flux ratio of the two images during the flare observed by the LAT in June 2012 \citep{marti13}.
This implies the presence of energy-dependent submillimeter structures in \pks\ during the $\gamma$-ray flare. While micro/milli-lensing events can already introduce a variability in the flux ratio, frequency-dependent changes directly imply an energy-dependent structure in the blazar nucleus like a ``core-shift'' effect (i.e. the frequency-dependent astrometric shift of the VLBI core position).
This discovery can have direct consequences for our observations considering the supposed mm/sub-mm and GeV $\gamma$-ray connection in blazars \citep{giommi12,marscher12}. The concurrence and co-spatiality of the submm and $\gamma$-ray June 2012 flares is a direct prediction of the shock-injet model, while the remarkable energy dependence of the flux ratio of the two mm/submm core images is related to opacity effects close to the base of the jet \citep{marti13}.

In radio bands dispersive refractive properties of the emitting plasma itself can cause
gravitational deflection angle to be dependent by the photon energy
\citep{bisnovatyi10}, but this effect can be considered not significant at GeV energies.
Another aspect is the presence of a strong
(cluster-scale) gravitational potential, even with strong lensing
only. Source emission anisotropy may create spectroscopic differences
along the slightly different lines of sight, yielding to differences in
relativistic beaming of the images and a certain probability that one
of the lensed image and delayed flare event may be not observable
\citep{perna09}. However, the Einstein angle is small for an isolated galaxy scale potential and consequently also source anisotropy is not significant in the
case of \pks.

The non-detection of delayed flares for the ``B'' and ``C'' $\gamma$-ray flares and the lack of correlated activity in soft X-rays observed by \swift\ do not interfere with the association and identification of this LAT source with the lensed background blazar \pks. This is because of the tighter spatial localization constraints towards \pks\ coming from the 1FGL, 2FGL (and the next 3FGL) Catalogs for the source. Additionally, the lensing galaxies located at $z=0.88582$ and $z=0.19$ are unlikely to be bright $\gamma$-ray sources, being a passive faint red galaxy and a passive face-on spiral galaxy \citep{courbin02,winn02} respectively. The identification of \pks\ as a $\gamma$-ray source is declared since the EGRET era \citep{mattox97a,combi98}.

Though initially considered a
simple two-image gravitational lens, the lensed $\gamma$-ray quasar \pks\ appears to have several peculiar and intriguing features. The line of sight to \pks\
appears to be very busy: one possible Galactic main-sequence star, and two or (more likely) three lensing galaxies \citep{courbin02}. \pks\ represents also the first known case of a quasar lensed by an almost face-on spiral galaxy \citep{courbin02,winn02}, where a different flux suppression for the two different lens image paths represents another hypothesis.

No lens model has been able to explain yet all the observed characteristics and physical phenomena associated with the lens galaxies and the background blazar. As example the same radio time delay value $\sim 26$ days could be replaced by a more secure range of possible time delays ranging from 12 to 30 days, based on the full set of light curves used by \citet{lovell98}.
There is also evidence for substructure in this lens and the true mass distribution os the system is probably more complicated than the distributions in published lens models for \pks\ \citep{jin03}.

Deep optical imaging of \pks\ does not
produce a clear picture of the lens and surrounding field because the line
of sight lies near to the Galactic plane and the bulge of the Milky Way. Modeling of \pks\ has not been able to derive the Hubble constant with the precision obtained using other cosmological lenses. Besides uncertainty in the measured radio-time delay \pks\ has also remarkable uncertainties in the localization of the lensing galaxy and lens barycenter.

The continuous all-sky survey monitoring performed in the next years by \fermi\ LAT during the extended mission era, and the future Pass 8 data release, based on a complete revision of the entire event-level analysis,
will allow the production of improved light curves for more detailed analysis. \pks\ may be the best high-energy
gravitational lens for simultaneous mm/sub-mm and $\gamma$-ray variability and lensing studies with ALMA and the \fermi\ LAT.

%
\section{Acknowledgments}
%
%
\footnotesize{We thank the anonymous referees for useful comments that improved the paper.
SC thanks Dr. R. Porcas of MPIfR, Bonn,
Germany for a useful discussion during the course of this work.
The \textit{Fermi} LAT Collaboration acknowledges generous ongoing support
from a number of agencies and institutes that have supported both the
development and the operation of the LAT as well as scientific data analysis.
These include the National Aeronautics and Space Administration and the
Department of Energy in the United States, the Commissariat \`a l'Energie Atomique
and the Centre National de la Recherche Scientifique / Institut National de Physique
Nucl\'eaire et de Physique des Particules in France, the Agenzia Spaziale Italiana
and the Istituto Nazionale di Fisica Nucleare in Italy, the Ministry of Education,
Culture, Sports, Science and Technology (MEXT), High Energy Accelerator Research
Organization (KEK) and Japan Aerospace Exploration Agency (JAXA) in Japan, and
the K.~A.~Wallenberg Foundation, the Swedish Research Council and the
Swedish National Space Board in Sweden.
\par Additional support for science analysis during the operations phase is gratefully
acknowledged from the Istituto Nazionale di Astrofisica in Italy and the Centre National d'\'Etudes Spatiales in France.
\par This work includes observations obtained with the NASA \textit{Swift}
$\gamma$-ray burst Explorer. \swift\ is a MIDEX Gamma Ray Burst mission led by NASA with participation of Italy and the UK.
\par This research has made use of  the Smithsonian/NASA's ADS bibliographic
database. This research has made use of the archives and services of the ASI Science Data Center (ASDC), a facility of the Italian Space Agency (ASI Headquarter, Rome, Italy). This research has made use of the NASA/IPAC NED database (JPL CalTech and NASA, USA).
}


{\em Facilities:} \facility{ {\em~ Fermi} LAT},  \facility{Swift}

\bibliographystyle{apj}

\end{document}